***********************************************************************
     
\font\twelverm=cmr10 scaled 1200    \font\twelvei=cmmi10 scaled 1200
\font\twelvesy=cmsy10 scaled 1200   \font\twelveex=cmex10 scaled 1200
\font\twelvebf=cmbx10 scaled 1200   \font\twelvesl=cmsl10 scaled 1200
\font\twelvett=cmtt10 scaled 1200   \font\twelveit=cmti10 scaled 1200
\font\twelvesc=cmcsc10 scaled 1200  %\font\twelvesf=cmssmc10 scaled 1200
\skewchar\twelvei='177   \skewchar\twelvesy='60
     
%  Define \...point macros to change fonts and spacings consistently
     
\def\twelvepoint{\normalbaselineskip=12.4pt plus 0.1pt minus 0.1pt
  \abovedisplayskip 12.4pt plus 3pt minus 9pt
  \belowdisplayskip 12.4pt plus 3pt minus 9pt
  \abovedisplayshortskip 0pt plus 3pt
  \belowdisplayshortskip 7.2pt plus 3pt minus 4pt
  \smallskipamount=3.6pt plus1.2pt minus1.2pt
  \medskipamount=7.2pt plus2.4pt minus2.4pt
  \bigskipamount=14.4pt plus4.8pt minus4.8pt
  \def\rm{\fam0\twelverm}          \def\it{\fam\itfam\twelveit}%
  \def\sl{\fam\slfam\twelvesl}     \def\bf{\fam\bffam\twelvebf}%
  \def\mit{\fam 1}                 \def\cal{\fam 2}%
  \def\sc{\twelvesc}               \def\tt{\twelvett}
  \def\sf{\twelvesf}
  \textfont0=\twelverm   \scriptfont0=\tenrm   \scriptscriptfont0=\sevenrm
  \textfont1=\twelvei    \scriptfont1=\teni    \scriptscriptfont1=\seveni
  \textfont2=\twelvesy   \scriptfont2=\tensy   \scriptscriptfont2=\sevensy
  \textfont3=\twelveex   \scriptfont3=\twelveex  \scriptscriptfont3=\twelveex
  \textfont\itfam=\twelveit
  \textfont\slfam=\twelvesl
  \textfont\bffam=\twelvebf \scriptfont\bffam=\tenbf
  \scriptscriptfont\bffam=\sevenbf
  \normalbaselines\rm}
     
%       tenpoint

%%
%%      Various internal macros
%%
     
\def\beginlinemode{\endmode
  \begingroup\parskip=0pt \obeylines\def\\{\par}\def\endmode{\par\endgroup}}
\def\beginparmode{\endmode
  \begingroup \def\endmode{\par\endgroup}}
\let\endmode=\par
{\obeylines\gdef\
{}}
\def\singlespace{\baselineskip=\normalbaselineskip}

\def\oneandahalfspace{\baselineskip=\normalbaselineskip
  \multiply\baselineskip by 3 \divide\baselineskip by 2}
\def\doublespace{\baselineskip=\normalbaselineskip \multiply\baselineskip by 2}

\newcount\firstpageno
\firstpageno=2
\footline={\ifnum\pageno<\firstpageno{\hfil}\else{\hfil\twelverm\folio\hfil}\fi}
\def\toppageno{\global\footline={\hfil}\global\headline
  ={\ifnum\pageno<\firstpageno{\hfil}\else{\hfil\twelverm\folio\hfil}\fi}}
\let\rawfootnote=\footnote              % We must set the footnote style
\def\footnote#1#2{{\rm\singlespace\parindent=0pt\parskip=0pt
  \rawfootnote{#1}{#2\hfill\vrule height 0pt depth 6pt width 0pt}}}
\def\raggedcenter{\leftskip=4em plus 12em \rightskip=\leftskip
  \parindent=0pt \parfillskip=0pt \spaceskip=.3333em \xspaceskip=.5em
  \pretolerance=9999 \tolerance=9999
  \hyphenpenalty=9999 \exhyphenpenalty=9999 }
\def\dateline{\rightline{\ifcase\month\or
  January\or February\or March\or April\or May\or June\or
  July\or August\or September\or October\or November\or December\fi
  \space\number\year}}
\def\received{\vskip 3pt plus 0.2fill
 \centerline{\sl (Received\space\ifcase\month\or
  January\or February\or March\or April\or May\or June\or
  July\or August\or September\or October\or November\or December\fi
  \qquad, \number\year)}}
     
%%
%%      Page layout, margins, font and spacing (feel free to change)
%%

\hsize=6.5truein
%\hoffset=1truein
\vsize=8.5truein
%\voffset=-1.0truein

\parskip=\medskipamount
\def\\{\cr}
\twelvepoint            % selects twelvepoint fonts (cf. \tenpoint)
\doublespace            % selects double spacing for main part of paper (cf.
                        %       \singlespace, \oneandahalfspace)
\overfullrule=0pt       % delete the nasty little black boxes for overfull box

\def\title                      %  Title on title page
  {\null\vskip 3pt plus 0.2fill
   \beginlinemode \doublespace \raggedcenter \bf}
     
\def\author                     %  Author(s) name(s)  on title page
  {\vskip 3pt plus 0.2fill \beginlinemode
   \singlespace \raggedcenter\sc}
     
\def\affil                      % Affiliations (can intermix with \author)
  {\vskip 3pt plus 0.1fill \beginlinemode
   \oneandahalfspace \raggedcenter \sl}
     
\def\abstract                   % Begin abstract
  {\vskip 3pt plus 0.3fill \beginparmode
   \singlespace ABSTRACT: }
     
\def\endtopmatter               % End title page, begin body of paper
  {\endpage                     %       This subsumes \body
   \body}
     
\def\body                       % Begin text body;  can be used to end
  {\beginparmode}               % \title, \author, \affil, \abstract,
                                % \reference, or \figurecaption modes
     
\def\head#1{                    % Head;  NOTE enclose the text in {}
  \goodbreak\vskip 0.5truein    %  e.g., \head{I. Introduction}
  {\immediate\write16{#1}
   \raggedcenter \uppercase{#1}\par}
   \nobreak\vskip 0.25truein\nobreak}
     
\def\subhead#1{                 % Subhead;  NOTE enclose the text in {}
  \vskip 0.25truein             % e.g., \subhead{A. History of the Problem}
  {\raggedcenter {#1} \par}
   \nobreak\vskip 0.25truein\nobreak}
     
\def\beginitems{
\par\medskip\bgroup\def\i##1 {\item{##1}}\def\ii##1 {\itemitem{##1}}
\leftskip=36pt\parskip=0pt}
\def\enditems{\par\egroup}
     
\def\beneathrel#1\under#2{\mathrel{\mathop{#2}\limits_{#1}}}
     
\def\refto#1{$^{#1}$}           % For references in text as superscript
     
\def\references                 % Begin references -- basic format is Phys Rev
  {\head{References}            % I.e., volume, page, year (space after commas).
   \beginparmode
   \frenchspacing \parindent=0pt \leftskip=1truecm
   \parskip=8pt plus 3pt \everypar{\hangindent=\parindent}}

\gdef\refis#1{\item{#1.\ }}                     % Ref list numbers.
     
\gdef\journal#1, #2, #3, 1#4#5#6{               % Journal reference.  Comma sets
    {\sl #1~}{\bf #2}, #3 (1#4#5#6)}            % off: name, vol, page, year

\gdef\refa#1, #2, #3, #4, 1#5#6#7.{\noindent#1, #2 {\bf #3}, #4 (1#5#6#7).\rm} 
%refa: type in: name, 
%journal, vol, page, year
%prints out in same order

\gdef\refb#1, #2, #3, #4, 1#5#6#7.{\noindent#1 (1#5#6#7), #2 {\bf #3}, #4.\rm} 
%refb: reads in same
%prints out name (year) etc.

\def\pr{\journal Phys.Rev., }

\def\prl{\journal Phys.Rev.Lett., }
     
\def\jmp{\journal J.Math.Phys., }
     
\def\rmp{\journal Rev.Mod.Phys., }

\def\np{\journal Nucl.Phys., }

\def\annp{\journal Ann.Phys.(N.Y.), }

\def\endreferences{\body}

\def\endpage                    %  Eject a page
  {\vfill\eject}
     
\def\endpaper                   %  Ways to say goodbye
  {\endmode\vfill\supereject}

\def\ref#1{Ref.~#1}                     %       for inline references
\def\Ref#1{Ref.~#1}                     %       ditto
\def\[#1]{[\cite{#1}]}
\def\cite#1{{#1}}
\def\(#1){(\call{#1})}
\def\call#1{{#1}}
\def\taghead#1{}
\def\frac#1#2{{#1 \over #2}}
\def\half{{\frac 12}}

\def\12{{1\over2}}

\catcode`@=11
\newcount\r@fcount \r@fcount=0
\newcount\r@fcurr
\immediate\newwrite\reffile
\newif\ifr@ffile\r@ffilefalse
\def\w@rnwrite#1{\ifr@ffile\immediate\write\reffile{#1}\fi\message{#1}}

\def\writer@f#1>>{}
\def\referencefile{%			  Stuff to write .REF file
  \r@ffiletrue\immediate\openout\reffile=\jobname.ref%
  \def\writer@f##1>>{\ifr@ffile\immediate\write\reffile%
    {\noexpand\refis{##1} = \csname r@fnum##1\endcsname = %
     \expandafter\expandafter\expandafter\strip@t\expandafter%
     \meaning\csname r@ftext\csname r@fnum##1\endcsname\endcsname}\fi}%
  \def\strip@t##1>>{}}

\def\citeall#1{\xdef#1##1{#1{\noexpand\cite{##1}}}}
\def\cite#1{\each@rg\citer@nge{#1}}	% Variable No. of args, separated by

\def\each@rg#1#2{{\let\thecsname=#1\expandafter\first@rg#2,\end,}}
\def\first@rg#1,{\thecsname{#1}\apply@rg}	% each@ag is a general purpose
\def\apply@rg#1,{\ifx\end#1\let\next=\relax%	  variable no. of arg. macro.
\else,\thecsname{#1}\let\next=\apply@rg\fi\next}% args separated by commas

\def\citer@nge#1{\citedor@nge#1-\end-}	% Check for M-N range (M and N numbers)
\def\citer@ngeat#1\end-{#1}
\def\citedor@nge#1-#2-{\ifx\end#2\r@featspace#1 % Single argument
  \else\citel@@p{#1}{#2}\citer@ngeat\fi}	% M-N range of arguments
\def\citel@@p#1#2{\ifnum#1>#2{\errmessage{Reference range #1-#2\space is bad.}%
    \errhelp{If you cite a series of references by the notation M-N, then M and
    N must be integers, and N must be greater than or equal to M.}}\else%
 {\count0=#1\count1=#2\advance\count1 by1\relax\expandafter\r@fcite\the\count0,
  \loop\advance\count0 by1\relax%	  Loop from M to N
    \ifnum\count0<\count1,\expandafter\r@fcite\the\count0,%
  \repeat}\fi}

\def\r@featspace#1#2 {\r@fcite#1#2,}	% Eat spaces at beginning or end of arg
\def\r@fcite#1,{\ifuncit@d{#1}%		  Cite individual reference
    \newr@f{#1}%
    \expandafter\gdef\csname r@ftext\number\r@fcount\endcsname%
                     {\message{Reference #1 to be supplied.}%
                      \writer@f#1>>#1 to be supplied.\par}%
 \fi%
 \csname r@fnum#1\endcsname}
\def\ifuncit@d#1{\expandafter\ifx\csname r@fnum#1\endcsname\relax}%
\def\newr@f#1{\global\advance\r@fcount by1%
    \expandafter\xdef\csname r@fnum#1\endcsname{\number\r@fcount}}

\let\r@fis=\refis			% Save old \refis, redefine
\def\refis#1#2#3\par{\ifuncit@d{#1}%      Use two params #2 #3 to strip blank
   \newr@f{#1}%
   \w@rnwrite{Reference #1=\number\r@fcount\space is not cited up to now.}\fi%
  \expandafter\gdef\csname r@ftext\csname r@fnum#1\endcsname\endcsname%
  {\writer@f#1>>#2#3\par}}

\def\ignoreuncited{%   redefine \refis if ignoring uncited references
   \def\refis##1##2##3\par{\ifuncit@d{##1}%
    \else\expandafter\gdef\csname r@ftext\csname r@fnum##1\endcsname\endcsname%
     {\writer@f##1>>##2##3\par}\fi}}

\def\r@ferr{\endreferences\errmessage{I was expecting to see
\noexpand\endreferences before now;  I have inserted it here.}}
\let\r@ferences=\references
\def\references{\r@ferences\def\endmode{\r@ferr\par\endgroup}}

\let\endr@ferences=\endreferences
\def\endreferences{\r@fcurr=0%		  Save old \endreferences, redefine
  {\loop\ifnum\r@fcurr<\r@fcount%	  Loop over refnum and produce text
    \advance\r@fcurr by 1\relax\expandafter\r@fis\expandafter{\number\r@fcurr}%
    \csname r@ftext\number\r@fcurr\endcsname%
  \repeat}\gdef\r@ferr{}\endr@ferences}

% Save old \endpaper, redefine it to write parting message.

\let\r@fend=\endpaper\gdef\endpaper{\ifr@ffile
\immediate\write16{Cross References written on []\jobname.REF.}\fi\r@fend}

\catcode`@=12

\citeall\refto		% These macros will generate citations
\citeall\ref		%
\citeall\Ref		%

\def\a{{\alpha}}
\def\b{{\beta}}
\def\c{{\gamma}}
\def\d{{\delta}}

\def\e{{\epsilon}}
\def\half{{1 \over 2}}
\def\ra{{\rangle}}
\def\la{{\langle}}

\def\ih{{i \over \hbar}}
\def\l{\ell}

\def\au{\underline \alpha}
\def\p{{\partial}}

\def\ria{{\rightarrow}}
\def\Tr{{\rm Tr}}

\centerline{\bf Decoherent Histories Approach to}
\centerline{\bf the Arrival Time Problem}

\author J.J.Halliwell % \footnote{$^{\dag}$}{E-mail address: \jjh}
\vskip 0.2in
\centerline{\rm and}
\author E.Zafiris
\affil
Theory Group
Blackett Laboratory
Imperial College
London SW7 2BZ
UK
\vskip 0.5in
\centerline {\rm Preprint IC 96-97/04. June, 1997}
\vskip 0.1in
\centerline {\rm quant-ph/9706045}
\vskip 0.1in 
\centerline {\rm Submitted to {\sl Physical Review D}}
\vskip 1.0in
\abstract

{What is the probability of a particle entering a given region of
space at {\it any} time between $t_1$ and $t_2$? Standard quantum
theory assigns probabilities to alternatives at a fixed moment of
time and is not immediately suited to questions of this type. We use
the decoherent histories approach to quantum theory to compute the
probability of a non-relativistic particle crossing $x=0$ during an
interval of time. For a system consisting of a single
non-relativistic particle, histories  coarse--grained according to
whether or not they pass through spacetime regions are generally not
decoherent, except for very special initial states, and thus
probabilities cannot be assigned. Decoherence may, however, be
achieved by coupling the particle to an environment consisting of a
set of harmonic oscillators in a thermal bath.  Probabilities for
spacetime coarse grainings are thus calculated by considering
restricted density operator propagators of the quantum Brownian
motion model. We also show how to achieve decoherence by replicating
the system $N$ times and then projecting onto the number density of
particles that cross during a given time interval, and this gives an
alternative expression for the crossing probability. The latter
approach shows that the relative frequency for histories is
approximately decoherent for sufficiently large $N$, a result
related to the Finkelstein-Graham-Hartle theorem.
}
\endtopmatter 
\endpage

\head{\bf 1. Introduction}

In non-relativistic quantum mechanics, the probability of finding a
particle between points $x $ and $ x + dx $ at a fixed time $ t $ is given
by
$$
p(x,t) dx = | \Psi (x,t ) |^2 \ dx
\eqno(1.1)
$$
where $\Psi( x,t)$ is the wave function of the particle. More
generally, the variety of questions one might ask about a particle
at a fixed moment of time may be represented by a projection
operator $P_{\a}$, and the probability of a particular alternative
is given by
$$
p(\a ) = \Tr \left( P_{\a} \rho \right)
\eqno(1.2)
$$
where $\rho$ is the density operator of the system.

Eqs.(1.1) and (1.2) refer to questions about the properties of the
particle at a fixed moment of time. However, it is of interest to
ask questions about the particle that do not refer to a particular
moment of time. One could ask, for example, for the probability that
the particle entered the region between $x$ and $x+dx $ at {\it any}
moment of time between $t_1 $ and $t_2$. That is, for the
probability of finding the particle in a region of {\it spacetime}. 
What predictions does quantum mechanics make for questions of this
type?

This question is clearly a physically relevant one since time is
measured by physical devices which are generally limited in their
precision. It is therefore never possible to say that a physical
event occurs at a precise value of time, only that it occurs in some
range of times. Furthermore, there has been considerable recent
experimental and theoretical interest in the question of tunneling
times [\cite{HaS,Lan}]. That is, the question, given that a particle
has tunneled through a barrier region, how much time did it spend
inside the barrier?

The question of time in non-relativistic quantum mechanics is also
closely related to the so-called ``problem of time'' in quantum
gravity. In quantum cosmology, the wave function of the
universe satisfies not a Schr\"odinger equation, but the
Wheeler--DeWitt equation, 
$$ 
{\cal H} \Psi [h_{ij}, \phi ] = 0 
\eqno(1.3)
$$
The wave function $\Psi$ depends on the three-metric $h_{ij}$ and
the matter field configurations $\phi$ on a closed  spacelike
three-surface [\cite{Har1,Har3,Hal4}].
There is no time label. ``Time'' is somehow already
present amongst the dynamical variables $h_{ij}, \phi $. Although a
comprehensive scheme for interpreting the wave function is yet to be
put forward, one possible view is that the
interpretation will involve treating all the dynamical variables
$h_{ij}, \phi$ on an equal footing, rather than trying to single out
one particular combination of them to act as time.  For this reason, it is of
interest to see if one can carry out a similar exercise in
non-relativistic quantum mechanics. That is, to  see what the
predictions quantum mechanics makes about {\it spacetime} regions,
rather than regions of space at fixed moments of time.

Spacetime questions tend to be rather non-trivial. As
stressed by Hartle, who has carried out a number of investigations
in this area [\cite{Har2,Har0,Har}],  time plays a ``peculiar and
central role'' in non-relativistic quantum mechanics. It is not
represented by a self-adjoint operator and there is no obstruction
to assuming that it may be measured with arbitrary precision. It
enters the Schr\"odinger equation as an external parameter. As such,
it is perhaps best thought of as a label referring to a classical,
external measuring device, rather than as a fundamental quantum
observable. Yet time is  measured by physical systems, and all
physical systems are believed to be subject to the laws of quantum
theory.

Given these features, means more elaborate that those usually
employed are required to define quantum-mechanical probabilities
that do not refer to a specific moment of time, and the issue has a
long history [\cite{Time}]. One may find in the literature a variety
of attempts to define questions of time in a quantum--mechanical
way. These include attempts to define time operators
[\cite{GRT,Hol,Per}], the use of internal physical clocks 
[\cite{Har2,Har0}] and path integral approaches
[\cite{Fer,YaT,Har,Kum}]. The literature on tunneling times is a
particularly rich source of ideas on this topic [\cite{HaS}].  Many
of these attempts also tie in with the time--energy uncertainty
relations [\cite{MaT,KoA}].

The approach we shall use in this paper involves the  decoherent
histories approach to quantum theory [\cite{GeH1,GeH2,Gri,Omn}].
This is an approach to quantum
theory suitable for genuinely closed systems. It was developed in
part for quantum cosmology, but it has been very fruitful in
enhancing understanding of non-relativistic quantum systems,
especially the emergence of classical behaviour.

For our purposes, the particular attraction of this approach is that
assigns probabilities directly to the possible histories of a
system, rather than to events at a single moment of time. It is
therefore very suited to the question of spacetime probabilities
considered here. This is because the question of whether a particle
did or did not enter a given region at {\it any} time between $t_1$
and $t_2$ clearly cannot be reduced to a question about the state of
the particle at a fixed moment of time, but depends on the entire
history of the system during that time interval.

The decoherent histories approach, for spacetime questions, turns
out to be most clearly formulated in terms of path integrals over
paths in configuration space [\cite{YaT,Har,Har3}].
The desired spacetime amplitudes are
obtained by summing $e^{iS[x(t)]}$, where $S[x(t)]$ is the action,
over paths $x(t)$ passing through the spacetime region in question,
and consistent with the initial state. The probabilities are
obtained by squaring the amplitudes in the usual way.  The
decoherent histories approach is not inextricably tied to path
integrals, however. Operator approaches to the same questions are
also available, but are often more cumbersome.

The decoherent histories approach brings a new element into the game
which, it is clear from the literature, has so far only been
partially appreciated. This new feature is decoherence -- the
destruction of interference between histories.

When computed according to the path integral scheme outlined above,
the probability of entering a spacetime region added to the
probability of not entering that region is not equal to $1$, in
general.
This is because of interference. The question of whether a
particle enters a spacetime region, when carefully broken down, is
actually a quite complicated combination of questions about the
positions of the particle at a sequence of times. It is therefore,
in essence, a complicated combination of double slit situations. Not
surprisingly, there is therefore interference and probabilities
cannot be assigned.

This feature has been exhibited very clearly by the extensive work
of Yamada and Takagi [\cite{YaT}]. They considered a number of spacetime
coarse grainings for a free, non-relativistic particle. They found
that probabilities could be assigned, in the decoherent histories
approach, only for very special initial states, and the
probabilities were then rather uninteresting, {\it e.g.},
probability zero for entering the region, and $1$ for not entering
it. 

There is an important lesson here. For a free, non-relativistic
particle, probabilities for whether or not the particle enters a
spacetime region {\it cannot be assigned in general}, due to the
presence of interference. Physically, this may at first appear
unreasonable, because one could imagine situating a measuring device
in the spatial region in question, and then asking whether it
registers the presence of a particle during a given time interval.
The point, of course, is that introducing a measuring device
modifies the physical situation. A measuring device typically has a
large number of internal degrees of freedom, and, from the point of
view of the decoherent histories approach, these provide an
``environment'' which produces the decoherence necessary for the
assignment of probabilities.  (This is in keeping with the general
point made by Landauer in the context of tunneling times -- that
time in quantum mechanics only makes sense if the mechanism by which
it is measured is fully specified [\cite{Lan}].)

Generally, therefore, we might expect that by making suitable
modifications to the basic physical situation, decoherence may be
achieved and probabilities may be assigned to spacetime coarse
grainings. In this paper, we will consider two simple modifications
which lead to decoherence for spacetime coarse grainings of a point
particle.

The first modification consists of coupling the point particle to a
bath of harmonic oscillators in a thermal state (the quantum
Brownian motion model [\cite{CaL,FeV}]). Interference is destroyed as a
result of the interaction with the bath, and probabilites can be
assigned for essentially arbitrary initial states of the point
particle. This modification is a model of continuous position
measurements. 

The second modification is to replicate the system $N$ times, where
$N$ is large, and then ask for the probability that some fraction of
the particles $f= n/N$ enters the spacetime region. Coarse graining
$f$ over a small range, together with large $N$ statistics then
ensures decoherence, and probabilities may then be assigned to $f$.
The reason we expect decoherence here is that we are effectively
projecting onto number density, which is expected to be decoherent
because it is typically a slowly varying quantity [\cite{BrH}].
This modification is less obviously tied to a particular type of
measurement, but it can be shown that decoherent sets of histories
correspond, in a certain sense, to {\it some} kind of measurement
(not necessarily a physically realizable one) [\cite{GeH2,GeH4}].

In Section II, we briefly review the decoherent histories approach.
In Section III, we briefly review the work of Hartle and of Yamada
and Takagi on spacetime coarse grainings. In Section IV we sketch
our results on spacetime coarse grainings for quantum Brownian
motion models. In Section V, we describe the large $N$ case.
We summarize and conclude in Section VI.

\head{\bf 2. Decoherent Histories Approach to Quantum Theory}

We give here a very brief summary of the decoherent histories
approach to quantum theory. Far more extensive descriptions can be
found in many other places 
[\cite{GeH1,GeH2,Gri,Hal1,Hal2,Har3,Ish,Omn,DoH}].

In quantum mechanics, propositions
about the attributes of a system at a fixed moment of time are
represented by sets of projections operators. The projection
operators $P_{\a}$ effect a partition of the possible alternatives
$\a$ a system may exhibit at each moment of time. They are
exhaustive and exclusive,
$$
\sum_{\a} P_{\a} =1, \quad \quad
P_{\a} P_{\beta} = \delta_{\a \beta} \ P_{\a}
\eqno(2.1)
$$
A projector is said to be {\it fine-grained} if it is of the form
$ | \a \ra \la \a | $, where $\{| \a \ra \}$ are a complete set of
states. Otherwise it is {\it coarse-grained}.
A quantum-mechanical history (strictly, a {\it homogeneous} history [\cite{Ish}])
is characterized by a string of time-dependent projections,
$P_{\a_1}^1(t_1), \cdots P_{\a_n}^n(t_n)$, together with an initial
state $\rho$. The time-dependent projections are related to the
time-independent ones by
$$
P^k_{\a_k}(t_k) = e^{i H(t_k-t_0)} P^k_{\a_k} e^{-i H(t_k-t_0)} 
\eqno(2.2)
$$
where $H$ is the Hamiltonian.
The candidate probability for these homogeneous histories is
$$
p(\a_1, \a_2, \cdots \a_n) = {\rm Tr} \left( P_{\a_n}^n(t_n)\cdots 
P_{\a_1}^1(t_1)
\rho P_{\a_1}^1 (t_1) \cdots P_{\a_n}^n (t_n) \right)
\eqno(2.3)
$$

It is straightforward to show that (2.3) is both non-negative and
normalized to unity when summed over $\a_1, \cdots \a_n$. 
However,
(2.3) does not satisfy all the axioms of probability theory, and for
that reason it is referred to as a candidate probability. It does
not satisfy the requirement
of additivity on disjoint regions of sample space. More precisely,
for each set of histories, one may construct coarser-grained
histories by grouping the histories together. This may be achieved,
for example, by summing over the projections at each moment of time,
$$
{\bar P}_{{\bar \a}} = \sum_{\a \in {\bar \a} } P_{\a}
\eqno(2.4)
$$
(although this is not the most general type of coarse graining
-- see below).
The additivity requirement is then that the probabilities for each
coarser-grained history should be the sum of the probabilities of
the finer-grained histories of which it is comprised.
Quantum-mechanical interference generally prevents this requirement
from being satisfied. Histories of closed quantum systems
cannot in general be assigned probabilities.

There are, however, certain types of histories for which
interference is negligible, and the candidate probabilities for histories
do satisfy the sum rules.
These histories may be found using the decoherence functional:
$$
D({\underline {\a}} , {\underline {\a}'} ) = 
\Tr \left( P_{\a_n}^n(t_n)\cdots 
P_{\a_1}^1(t_1)
\rho P_{\a_1'}^1 (t_1) \cdots P_{\a_n'}^n (t_n) \right)
\eqno(2.5)
$$
Here $ {\underline {\a}} $ denotes the string $\a_1, \a_2, \cdots
\a_n$. Intuitively, the decoherence functional measures the amount
of interference between pairs of histories.
It may be shown that the additivity requirement is satisfied for all
coarse-grainings if and only if 
$$
{\rm Re} D({\underline {\a}} , {\underline {\a}'} ) = 0 
\eqno(2.6)
$$
for all distinct pairs of
histories ${\underline {\a}}, {\underline {\a}'}$ [\cite{Gri}].
Such sets of histories are said to be {\it consistent}, or 
{\it weakly decoherent.} 
The consistency condition (2.6) is typically satisfied only
for coarse--grained histories, and this then often leads
to satisfaction of the stronger condition of {\it decoherence}
$$
D(\au, \au' ) = 0
\eqno(2.7)
$$
for $\au \ne \au' $.
The condition of decoherence is associated with the existence of
so-called generalized records, corresponding to the idea that
information about the variables followed is stored in the variables
ignored in the coarse graining procedure [\cite{GeH2,GeH4}].

For histories characterized by projections onto ranges of position
at different times, the decoherence functional may be represented
by a path integral:
$$
D(\a, \a') = \int_{\a} {\cal D} x \ \int_{\a'} {\cal D} y 
\ \exp \left( \ih S[x] - \ih S[y] \right) \ \rho (x_0, y_0)
\eqno(2.8)
$$
The integral is over paths $x(t)$, $y(t)$ starting at $x_0$, $y_0$,
and both ending at the same final point $x_f$, where $x_f$, $x_0$
and $y_0$ are all integrated over, and weighted by the initial
state $\rho(x_0,y_0)$. The paths are also constrained to pass
through spatial gates at a sequence of times corresponding to the 
projection operators.

However, the path integral representation of the decoherence
functional also points the way towards asking types of questions
that are not represented by homogeneous histories [\cite{Har}]. Consider for
example the following question. Suppose a particle starts at $t=0$
in a state with non-zero support only in $x>0$. What is the
probability that the particle will either cross or never cross $x=0$
during the time interval $[0,\tau]$? In the path integral of the
form (2.8) it is clear
how to proceed. One sums over paths that, respectively,  either
always cross or never cross $x=0$ during the time interval. 

How does this look in operator language?
The operator form of the decoherence functional is 
$$
D(\a, \a') = \Tr \left( C_{\a} \rho C_{\a'}^{\dag} \right)
\eqno(2.9)
$$
where 
$$
C_{\a} = P_{\a_n}(t_n) \cdots P_{\a_1} (t_1)
\eqno(2.10)
$$
The histories that never cross $x=0$
are represented by taking the projectors in $C_{\a}$
to be onto the positive $x$-axis, and then taking
the limit $ n \ria \infty$ and $t_k - t_{k-1} \ria 0 $.
The histories that always cross $x=0$ are then represented by
the object
$$
{\bar C}_{\a} = 1 - C_{\a}
\eqno(2.11)
$$
This is called an {\it inhomogenous} history, because it cannot be
represented as a single string of projectors.  It can however, be
represented as a {\it sum} of strings of projectors [\cite{Har,Ish}].

The proper framework in which these operations, in particular (2.11),
are understood, is the so-called generalized quantum theory of
Hartle [\cite{Har}] and Isham {\it et al.} [\cite{Ish}]. 
It is called ``generalized''
because it admits inhomogeneous histories as viable objects, whilst
standard quantum theory concerns itself entirely with homogeneous
histories. We will make essential use of inhomogeneous histories
in what follows.

In practice, for point particle systems,
decoherence is readily achieved by coupling to an environment.
Here, we will use the much studied case of the quantum Brownian
motion model, in which the particle is linearly coupled through
position to a bath of harmonic oscillators in a thermal state
at temperature $T$ and characterized by a dissipation coefficient
$\gamma$. The details of this model may be found elsewhere 
[\cite{CaL,FeV,Hal1,Hal2}].

We consider histories characterized only by the position of the
particle and the environmental coordinates are traced out.
The path integral representation of the decoherence functional
then has the form
$$
D(\a, \a') = \int_{\a} {\cal D} x \ \int_{\a'} {\cal D} y 
\ \exp \left( \ih S[x] - \ih S[y] + \ih W[x,y] 
\right) \ \rho (x_0, y_0)
\eqno(2.12)
$$
where $W[x,y]$ is the Feynman--Vernon influence functional phase,
and is given by 
$$
W[x,y] = - m \gamma \int dt \ (x-y) ( \dot x + \dot y )
+ i { 2 m \gamma k T \over \hbar} \int dt \ (x-y)^2
\eqno(2.13)
$$
The first term induces dissipation in the effective classical
equations of motion. The second term is reponsible for thermal
fluctuations. It is also reponsible for suppressing contributions
from paths $x(t)$ and $y(t)$ that differ widely, and produces
decoherence of configuration space histories.

The corresponding classical theory is no longer the mechanics of a
single point particle, but a point particle coupled to a heat bath.
The classical correspondence is now to a stochastic process which
may be described by either a Langevin equation, or by a
Fokker-Planck equation for a phase space probability distribution
$w(p,x,t)$:
$$
{ \p w \over \p t}
= - { p \over m } { \p w \over \p x }
+ 2 \c { \p ( p w ) \over \p p }
+ D { \p^2 w \over \p p^2 }
\eqno(2.14)
$$
where $w \ge 0 $ and
$$
\int dp \ \int dx \ w(p,x,t) = 1
$$
When the mass is sufficiently large, this equation describes
near--deterministic evolution with small thermal fluctuations about it.

\head{\bf 3. Spacetime Coarse Grainings}

We are generally interested in spacetime coarse grainings which
consist of asking for the probability that a particle does or does
not enter a certain region of space during a certain time interval.
However, the essentials of this question boil down to
the following simpler question: what is the probability
that the particle will either cross or not cross $x=0$ at any time
in the time interval $[0,t]$? We will concentrate on this question.

In this section we briefly review the result of Yamada and Takagi
[\cite{YaT}], Hartle [\cite{Har,Har2,Har3}]  
and Micanek and Hartle [\cite{MiH}].
We will compute the decoherence functional using the path integral
expression (2.8), which may be written
$$
D(\a,\a') = \int dx_f \ \Psi^{\a}_t (x_f) \ \left( \Psi^{\a'}_t (x_f)
\right)^*
\eqno(3.1)
$$
where $\Psi^{\a}_t (x_f) $ denotes the amplitude obtained by summing
over paths ending at $x_f$ at time $t$,
consistent with the restriction $\a$ and consistent with
the given initial state, so we have
$$
\Psi^{\a}_t (x_f) = \int_{\a}
{\cal D} x(t) \ \exp \left( \ih S[x] \right) \ \Psi_0 (x_0)
\eqno(3.2)
$$

Suppose the system starts out in the initial state $\Psi_0 (x) $
at $t=0$. 
The amplitude for the particle to start in this initial state, and
end up at $x$ at time $t$, but without ever crossing
$ x = 0$, is
$$
\Psi_t^r (x) = \int_{-\infty}^{\infty} dx_0 \ g_r (x,t | x_0, 0 ) \ \Psi_0 (x_0)
\eqno(3.3)
$$
where $ g_r $ is the restricted Green function, {\it i.e.}, the sum
over paths that never crosses $x=0$. For the free particle
considered here (and also for any system with a potential symmetric
about $x=0$), $g_r$ may be constructed by the method of images:
$$
g_r (x,t|x_0,0) = \left( \theta (x) \ \theta (x_0 ) 
+ \theta (- x) \theta (- x_0) \right)
\ \left(g(x,t|x_0,0) - g(x,t|-x_0,0) \right)
\eqno(3.4)
$$
where $g(x,t|x_0,0)$ is the unrestricted propagator.

The amplitude to cross $x=0$ is 
$$
\Psi_t^c (x) = \int_{-\infty}^{\infty} dx_0 \ g_c (x,t | x_0, 0 ) \ \Psi_0 (x_0)
\eqno(3.5)
$$
where $g_c (x,t | x_0, 0 ) $ is the crossing propagator, {\it i.e.},
the sum over paths which always cross $x=0$. This breaks up into two
parts. If $x$ and $x_0$ are on opposite sides of $x=0$, it is
clearly just the usual propagator $ g(x,t| x_0, 0 ) $. If $x$ and
$x_0$ are on the same side of $x=0$, it is given by 
$ g (-x,t | x_0, 0) $. This may be seen by reflecting the
segment of the path after last crossing about $x=0$ [\cite{HaO}].
(Alternatively, this is just the usual propagator minus the
restricted one). Hence,
$$
\eqalignno{
g_c (x,t | x_0, 0 ) =& 
\left( \theta (x) \theta (-x_0 ) + \theta (-x) \theta ( x_0 )
\right) \ g (x,t | x_0, 0 )
\cr & +
\left( \theta (x) \ \theta (x_0 ) 
+ \theta (- x) \theta (- x_0) \right) \ g (-x, t | x_0, 0 )
&(3.6) \cr }
$$
The crossing propagator may also be expressed in terms of the
so-called
path decomposition expansion, a form which is sometimes useful
[\cite{AuK,vB,Hal3,HaO,ScZ}].

Inserting these expressions in the decoherence function, Yamada and
Takagi found that the consistency condition may be satisfied exactly
by states which are antisymmetric about $x=0$. The probability of
crossing $x=0$ is then $0$ and the probability of not crossing is
$1$. What is happening in this case is that the probability flux
across $ x=0 $, which clearly has non-zero components going both to
the left and the right, averages to zero. 

Less trivial probabilities are obtained in the case where one asks
for the probability that the particle remains always in $x>0$ or
not, with an initial state with support along the entire $x$-axis 
[\cite{Har3}]. The probabilities become trivial again, however, in
the interesting case of an initial state with support only in $x>0$.

Yamada and Takagi have also considered the case of the probability
of finding the particle in a spacetime region [\cite{YaT}]. That is, the
probability that the particle enters, or does not enter, the spatial
interval $\Delta$, at any time during the time interval $ [0,t]$.
Again the consistency condition is satisfied only for very special
initial states and the probabilities are then rather trivial.

In an attempt to assign probabilities for arbitrary initial states,
Micanek and Hartle considered the above
results in the limit that the time interval $[0,t]$ becomes very
small [\cite{MiH}]. Such an assignment must clearly be possible in
the limit $t \ria 0 $.
They found that both the off-diagonal terms of
the decoherence functional $D$ and the crossing probability $p$
are of order
$ \epsilon = \left( \hbar t / m \right)^{\half} $ for small $t$, and
the probability $\bar p$ for not crossing is of order $1$.
Hence $p + \bar p \approx 1 $. 
They therefore argued that probabilities can be assigned if $t$ 
is sufficiently small.

On the other hand, we have the exact relation,
$$ 
p + \bar p + 2 {\rm Re} D = 1
\eqno(3.7)
$$
$ {\rm Re} D $ represents the degree of fuzziness in the definition of
the probabilities. Since it is of the same order as $\bar p$, one may
wonder whether it is then valid to claim approximate consistency.
Another condition that may be relevant is the condition
$$ 
|D|^2 < < p \bar p 
\eqno(3.8) 
$$
which was suggested in Ref. [\cite{DoH}] as a measure of 
approximate decoherence, and is clearly satisfied in this case.
Ultimately, the question of which mathematical conditions best 
characterize approximate decoherence or approximate consistency can
only be settled by examining the means by which the predicted
probabilites could be tested experimentally, and this has not yet
been considered.

For a system consisting of a single point particle, therefore,
crossing probabilities can be assigned to histories only in a
limited class of circumstances. In the following sections, we will
see how probabilites may be assigned in a wider variety of
situations.

\head{\bf 4. Decoherence of Spacetime Coarse-Grained Histories
in the Quantum Brownian Motion Model}

In order to achieve decoherence for a wide class of initial states,
and hence assign probabilities to quantum-mechanical histories for
spacetime regions, it is necessary to modify the point particle
system in some way. In this section, we discuss a modification
consisting of coupling the particle to a bath of harmonic
oscillators in a thermal state. We are therefore considering the
quantum Brownian motion model, a model that has been discussed very
extensively in the decoherence literature [\cite{CaL}].

This explicit modification of the single particle system means that
the corresponding classical problem (to which the quantum results
should reduce under certain circumstances) is in fact a stochastic
process described by either a Langevin equation or by a
Fokker-Planck equation. It is therefore appropriate to first study
the arrival problem in the corresponding classical stochastic
process (see for example, Refs.[\cite{Sie,BuT,BoD,MaW,Zaf}], 
and references therein). 

\subhead{\bf 4(A). The Arrival Time Problem in Classical Brownian Motion}

Classical Brownian motion may be described by the Fokker-Planck
equation (2.14) for the phase space probability distribution
$w(p,x,t)$. For simplicity we will work in the limit of negligible
dissipation,
hence the equation is,
$$
{ \partial w \over \partial t} = - { p \over m }
{ \partial w \over \partial x} + D
{ \partial^2 w \over \partial p^2 }
\eqno(4.1)
$$
where $ D = 2 m \gamma k T $.
The Fokker-Planck equation is to be solved subject
to the initial condition
$$
w(p,x,0) = w_0 (p,x)
\eqno(4.2)
$$

Consider now the arrival time problem in classical Brownian
motion. The question is this. Suppose the initial state is localized
in the region $x>0$. What is the probability that, under evolution
according to the Fokker-Planck equation (4.1), the particle either
crosses or does not cross $x=0$ during the time interval $[0,t]$?

A useful way to formulate spacetime questions of this type
is in terms of the Fokker-Planck propagator,
$ K(p,x,t|p_0, x_0, 0 ) $. 
The solution to (4.1) with the initial condition
(4.2) may be written in terms of $K$ as,
$$
w (p,x,t) = 
\int_{- \infty}^{\infty} dp \int_{- \infty}^{\infty} dx
\ K(p,x,t|p_0,x_0,0) \ w_0 (p,x)
\eqno(4.3)
$$
The Fokker-Planck propagator satisfies the Fokker-Planck equation
(4.1) with respect to its final arguments, and satisfies 
delta function initial conditions,
$$
K(p,x,0|p_0,x_0,0) \ = \ \d (p-p_0) \ \d (x - x_0)
\eqno(4.4)
$$
For the free particle without dissipation, 
it is given explicitly by
$$
\eqalignno{
K(p,x,t|p_0,x_0,0) = N
\ \exp & \left( - \a (p - p_0 )^2 -\b (x- x_0 - {p_0 t \over  m} )^2 
\right.
\cr
& \quad \quad \quad \left. + \epsilon (p-p_0 )(x-x_0 - 
{p_0 t \over  m}) \right)
&(4.5) \cr }
$$
where $N$, $\a$, $\b$ and $\epsilon$ are given by
$$
\a = { 1 \over D t}, \quad \b = { 3m^2 \over Dt^3},
\quad \epsilon = { 3 m \over D t^2}, 
\quad N = \left( { 3 m^2 \over 4 \pi D^2 t^4} \right)^{\half}
\eqno(4.6)
$$
(with $ D = 2 m \gamma k T $).
An important property it satisfies is the composition law
$$
K (p,x,t |p_0,x_0,0) = \int_{-\infty}^{\infty} dp_1 \int_{-\infty}^{\infty}
dx_1 \ K(p,x,t|p_1,x_1,t_1) \ K(p_1,x_1,t_1 | p_0,x_0, 0 )
\eqno(4.7)
$$
where $ t > t_1 > 0 $.

For our purposes, the utility of the Fokker-Planck propagator is
that it may be used to assign probabilities to individual paths in
phase space. Divide the time interval
$[0,t]$ into subintervals, $t_0 = 0,t_1,t_2, \cdots t_{n-1}, t_n = t $.
Then in the limit that the subintervals go to zero, and $n \ria \infty $
but with $t$ held constant, the quantity
$$
\prod_{k=1}^n \ K(p_k, x_k, t_k | p_{k-1}, x_{k-1}, t_{k-1} )
\eqno(4.8)
$$
is proportional to
the probability for a path in phase space. The probability for
various types of coarse grained paths (including spacetime coarse
grainings) can therefore be calculated by summing over this basic
object. 

We are interested in the probability $w_r (p_n, x_n, t ) $ 
that the particle follows a path
which remains always in the region $ x> 0$ during the time interval
$ [0,t]$ and ends at the point $x_n > 0 $ with momentum $p_n$.
The desired total probabilities for crossing or not crossing can
then be constructed from this object. $w_r $ 
is clearly given by 
$$
\eqalignno{
w_r (p_n, x_n,t) = & 
\int_0^{\infty} dx_{n-1} \cdots \int_0^{\infty} dx_1 \int_0^{\infty}
dx_0 \int_{-\infty}^{\infty} dp_{n-1} \cdots 
\int_{-\infty}^{\infty} dp_1
\int_{-\infty}^{\infty} dp_0
\
\cr
& \times \ \prod_{k=1}^n \ K(p_k, x_k, t_k | p_{k-1}, x_{k-1}, t_{k-1} )
\ w_0 (p_0, x_0 )
&(4.9) \cr}
$$
in the continuum limit.

Now it is actually more useful to derive a differential equation and
boundary conditions for $w_r (p,x,t)$, rather than attempt to
evaluate the above multiple integral. First of all, it is clear 
from the properties of the propagator that
$ w_r (p,x,t) $ satisfies the Fokker-Planck equation (4.1) and 
the initial condition (4.2). However, we also expect some sort of
condition at $x=0$. From the explicit expression for the propagator
(4.5), (4.6), we see that in the continuum limit, the propagator between
$ p_{n-1}, x_{n-1} $ and the final point $p_n, x_n $ becomes proportional
to the delta function
$$
\delta \left( x_n  -x_{n-1} - p_n t / m  \right)
\eqno(4.10)
$$
Since $x_{n-1} \ge 0 $, when $x_n= 0$ this delta function will give
zero when $p_n > 0 $, but could be non-zero when $p_n < 0 $. Hence
we deduce that the boundary condition on $w_r (p,x,t) $ is
$$
w_r (p,0,t) = 0, \quad {\rm if } \quad p> 0 
\eqno(4.11)
$$
This is the absorbing boundary condition usually given for the
arrival time problem [\cite{MaW,WaU}] (although this argument for 
it does not seem to have appeared elsewhere).

It is now convenient to
introduce a restricted propagator $K_r (p,x,t|p_0,x_0,0)$, which 
propagates $w_r (p,x,t) $. That is,
$K_r $ satisfies the delta function initial conditions (4.4) and the same
boundary conditions as $w_r$, Eq.(4.11). Since the original Fokker-Planck
equation is not invariant under $x \rightarrow - x$, we cannot
expect that a simple method of images (of the type used in Section
3), will readily yield the restricted propagator $K_r$. 
$K_r$ has recently been found [\cite{BoD}], 
using a modified
method of images technique due to Carslaw [\cite{Car}], and we
briefly summarize those results.

Consider first the usual Fokker-Planck propagator (4.5).
Introducing the coordinates
$$
\eqalignno{
X & = {p \over m } - { 3 x \over 2 t }, \quad Y = { \sqrt {3} x \over
2 t}
&(4.12) \cr
X_0 & = - { p_0 \over 2 m } - { 3 x_0 \over 2 t},  
\quad Y_0 =  { \sqrt {3} \over 2 } \left( { p_0 \over m }
+ { x_0 \over t } \right) 
&(4.13) \cr }
$$
the propagator (4.5) becomes,
$$
K = { \sqrt {3} \over 2 \pi \tilde t^2 }
\ \exp \left( - { (X- X_0 )^2 \over \tilde t } 
- { (Y - Y_0 )^2 \over \tilde t } \right)
\eqno(4.14)
$$
Here, $\tilde t = D t / m^2 $.
Now go to polar coordinates,
$$
\eqalignno{
X & = r \cos \theta, \quad Y = r \sin \theta
&(4.15) \cr
X_0 & = r' \cos \theta', \quad Y_0 = r' \sin \theta' 
&(4.16) \cr }
$$
Then from (4.14), it is possible to construct a so-called multiform
Green function [\cite{Car}],
$$
g (r, \theta, r', \theta' ) =
{ \sqrt {3} \over 2 \pi^{3/2} \tilde t^2 }
\exp \left( - { r^2 + r'^2 - 2 r r' \cos (\theta - \theta')
\over \tilde t } \right) \ \int_{-\infty}^a d \lambda  \ e^{- \lambda^2}
\eqno(4.17)
$$
where 
$$
a = 2 \left( { r r' \over \tilde t } \right)^{\half} 
\ \cos \left( { \theta - \theta' \over 2} \right)
\eqno(4.18)
$$
Like the original Fokker-Planck propagator, this object is a
solution to the Fokker-Planck equation with delta function initial
conditions, but differs in that it has the property that it is
defined on a two-sheeted Riemann surface and has period $ 4 \pi $.
The desired restricted propagator $K_r$ is then given by
$$
K_r (p,x,t|p_0, x_0, 0 ) = g (r, \theta, r', \theta' )  -
g (r, \theta, r', - \theta' ) 
\eqno(4.19)
$$
The point $x=0$ for $p>0$ is $\theta = 0 $ in the new coordinates,
and the above object indeed vanishes at $ \theta = 0$. Furthermore,
the second term in the above goes to zero at $t=0$, whilst the
first one goes to a delta function as required.

The probability of not crossing the surface during the time interval
$ [0,t]$ is then given by
$$
p_r = \int_{-\infty}^{\infty} dp \int_0^{\infty} dx 
\int_{-\infty}^{\infty} dp_0 \int_0^{\infty} dx_0 
\ K_r (p,x,t | p_0, x_0, 0 ) \ w_0 ( p_0, x_0 )
\eqno(4.20)
$$
The probability of crossing must then be $p_c = 1 - p_r $,
which can also be written,
$$
p_c = \int_{-\infty}^0 dp 
\int_{-\infty}^{\infty} dp_0 \int_0^{\infty} dx_0 
\ { p \over m } \ K_r (p,x=0,t | p_0, x_0, 0 ) \ w_0 ( p_0, x_0 )
\eqno(4.21)
$$
This completes the discussion of the classical stochastic problem.

\subhead{\bf 4(B). The Arrival Time Problem in Quantum Brownian Motion}

We now consider the analagous problem in the quantum case.
We therefore attempt to repeat the analysis of Section 3, but using
instead of (3.1), the decoherence function
$$
D(\a, \a' ) =
\int_{\a} {\cal D} x \int_{\a'} {\cal D} y \ \exp \left( \ih S[x] 
- \ih S[y] + \ih W[x,y] \right) \ \rho_0 (x_0, y_0)
\eqno(4.22)
$$
Here, $W[x,y]$ is the influence funtional phase (2.13), but with the
dissipation term neglected.
The sum is over all paths $x(t)$, $y(t)$ which are consistent with
the coarse graining $\a$, $\a'$. Hartle has discussed how this case
might be carried out, and we follow his discussion [\cite{Har}].

Let the initial density operator have
support only on the positive axis, and we ask for the probability
that the particle either crosses or never crosses $x=0$ during the
time interval $[0,t]$. The history label $\a$ takes two values,
which we denote $\a=c$ and $\a=r$.
The decoherence functional is conveniently rewritten,
$$
D(\a, \a' ) = {\rm Tr} \left( \rho_{\a \a'} \right)
\eqno(4.23)
$$ 
where
$$
\eqalignno{
\la x_f | \rho_{\a \a'} | y_f \ra &\equiv \rho_{\a \a'} (x_f, y_f)
\cr
&= \int_{\a} {\cal D} x \int_{\a'} {\cal D} y \ \exp \left( \ih S[x] 
- \ih S[y] + \ih W[x,y] \right) \ \rho_0 (x_0, y_0)
&(4.24) \cr }
$$
where here the sum is over paths consistent with the coarse graining,
but they end at fixed final points $x_f$, $y_f$. This object
actually obeys a master equation,
$$
i \hbar { \partial  \rho \over \partial t} = 
- { \hbar^2 \over 2m } \left( { \partial^2 \rho \over \partial x^2 }
- { \partial^2 \rho \over \partial y^2 } \right)
- \ih D (x-y)^2 \rho 
\eqno(4.25)
$$
This is the usual master equation for the evolution of the density
operator of quantum Brownian motion [\cite{CaL}].

The objects $\rho_{\a \a'}$ are then found by solving this equation
subject to matching the initial state $\rho_0$, and also to
the following boundary conditions (which follow from the path
integral representation):
$$
\eqalignno{
\rho_{rr} (x,y) &= 0, \quad {\rm for} \quad x \le 0 \quad {\rm and}
\quad y \le 0 
&(4.26) \cr
\rho_{rc} (x,y) &= 0, \quad {\rm for} \quad x \le 0
&(4.27) \cr
\rho_{cr} (x,y) &= 0, \quad {\rm for} \quad y \le 0
&(4.28) \cr }
$$
Given $\rho_{rr}, \rho_{rc}, \rho_{cr} $, the quantity $\rho_{cc}$
may be calculated from the relation,
$$
\rho_{rr} + \rho_{rc} + \rho_{cr} + \rho_{cc} = \rho
\eqno(4.29)
$$

In the unitary case, this problem was solved very easily using the
method of images. From the results of Section 3, for example, it can
be seen that in the unitary case
$$
\rho_{rr} (x,y) = \theta (x) \theta (y)
\left( \rho (x,y) - \rho (-x,y) - \rho (x,-y) + \rho (-x,-y) \right)
\eqno(4.30)
$$
where $\rho (x,y) $ is the unrestricted solution to the master
equation matching the prescribed initial condition.

The problem in the non-unitary case treated here, however, is that
the master equation is {\it not} invariant under $x \ria -x $
(or under $ y \ria - y$), hence $\rho (-x,y)$ and $\rho(x,-y)$
are {\it not} solutions to the master equation. The method of images
is therefore not applicable in this case (contrary to the claim in
Ref.[\cite{Har}]). As far as an analytic
approach goes, this represent a very serious technical problem.
Restricted propagation problems are very hard to solve analytically
in the absence of the method of images. 

We will pursue an approximate analytic solution to the problem.
First, we will make use of the well-known fact that  evolution
according to the master equation (4.25) forces every initial density
operator to become approximately diagonal in position on a very
short time scale [\cite{JoZ,Zur,PHZ,PZ}]. 
(More generally, the density operator
approaches a form which is approximately diagonal in a set of phase
space localized states [\cite{HaZ}]). Therefore, all density
operators will satisfy the condition $\rho(x,0) = 0 = \rho (0,y) $
approximately, except perhaps when $x$ and $y$ are close to zero. So for
$\rho_{rc}$ and $\rho_{cr}$ the only solution satisfying the
boundary conditions (4.26), (4.27), in this approximation, is 
$$
\rho_{rc} (x,y) \approx 0, \quad \rho_{cr} (x,y) \approx 0
\eqno(4.31)
$$
This is actually not surprising, since these terms represent
the interference between histories, and the mechanism that makes
the density operator diagonal is also known to strongly suppress
interference between histories.
For $\rho_{rr}$ we still need to satisfy the boundary condition
(4.25) close to $x=y$. 

To proceed further, we make use of the Wigner representation of the
density operator [\cite{BaJ}]:
$$
\eqalignno{
W(p,x) &= { 1 \over 2 \pi \hbar} \int_{-\infty}^{\infty} 
d \xi \ e^{- \ih p \xi}
\ \rho( x + { \xi \over 2} , x - { \xi \over 2} )
&(4.32) \cr
\rho(x,y) &= \int_{-\infty}^{\infty} dp \ e^{\ih p (x-y) }
W (p, { x+y \over 2} )
&(4.33) \cr }
$$
The Wigner representation is very useful in studies of the
master equation, since it is similar to a classical phase
space distribution function. Indeed, for quantum Brownian 
motion model with a free particle, the Wigner function obeys
the same Fokker-Planck equation (4.1) as the analagous 
classical phase space distribution function. What makes
it fail to be a classical phase space distribution is that
it can take negative values. However, it can be shown that
the Wigner function becomes positive after a short time
(typically the decoherence time), and numerous authors
have discussed its use as an approximate classical phase
space distribution, under these conditions.

These properties suggest
that we can get an approximate solution to the quantum problem
by taking the solution to the classical stochastic problem
$w_r (p,x,t) $, and regarding it as the Wigner function of the
density operator. The desired density operator is then obtained
from the Wigner transform (4.33). 
The main issue is to demonstrate the connection between the quantum
and classical boundary conditions, (4.26) and (4.11).

The quantity $\rho_{rr} (x_f, y_f) $
is given by the path integral expression,
$$
\rho_{rr} (x_f, y_f)
= \int_r {\cal D} x \int_r {\cal D} y \ 
\exp \left( {i m \over 2 \hbar } \int dt \left( \dot x^2 - \dot y^2 \right)
- { 2 m \gamma k T \over \hbar^2 } \int dt (x-y)^2
\right) \ \rho_0 (x_0, y_0)
\eqno(4.34)
$$
where the subscript $ r $ denotes the fact that the path integral is
over paths $ x(t), y(t) $ that lie in $ x> 0, y> 0 $. 
This path integral is the exact solution to the master equation
(4.25) and the boundary conditions (4.26).
Now introduce
$ X = \half ( x + y ) $, $ \xi = x - y $. Then the path integral
becomes
$$
\eqalignno{
\rho_{rr} (x_f, y_f)
= \int_r {\cal D} X \int_r {\cal D} \xi \ 
& \exp \left( - {i m \over \hbar } \int dt   \xi \ddot X  
+ { i m \over \hbar } \xi_f \dot X_f
- { i m \over \hbar } \xi_0 \dot X_0
- { 2 m \gamma k T \over \hbar^2 } \int dt \xi^2
\right) 
\cr
& \times \ \rho_0 (X_0 + \half \xi_0, X_0 - \half \xi_0)
&(4.35) \cr }
$$
where an integration by parts has been performed in the exponent.

The ranges of integration of $X$ and $\xi$ are now unfortunately not
so simple. The region $ x>0, y> 0 $ translates into $ X > 0 $,
$ - X < \xi < X $. However, we may exploit the fact that the
constant $ 2 m \gamma k T / \hbar^2 $ is typically very large (it is
this that gives decoherence) so the integral over $\xi$ is strongly
concentrated around $\xi = 0$. This means that the range of $\xi$
may be extended to $ ( -\infty, \infty) $ and the Gaussian integral
over $\xi $ may be carried out. Furthermore, the integral over
$\xi_0$ effectively performs a Wigner transform of the initial
state, and we obtain,
$$
\rho_{rr} (x_f, y_f) = 
\int_r {\cal D} X \ \exp \left( { i m \over \hbar } \xi_f \dot X_f
- { m \over 8 \gamma k T}
\int dt \ddot X^2 \right) \ W_0 (m \dot X_0, X_0 ) 
\eqno(4.36)
$$
where in the functional integral over $X$, $X_f $ is fixed.
(A similar trick was used in Ref.[\cite{GeH2}]). 
Denoting the Wigner transform
of $\rho_{rr}$ by $W_{rr} $, this equation is readily rewritten as
$$
W_{rr} ( m \dot X_f, X_f ) =
\int_r {\cal D} X \ \exp \left( 
- { m \over 8 \gamma k T}
\int dt \ddot X^2 \right) \ W_0 (m \dot X_0, X_0 ) 
\eqno(4.37)
$$
where the functional integral over $X(t)$ is over paths which lie in
$X>0$, and match $X_f$ and $\dot X_f $ at the final time.

Now the point is that the path integral (4.37) is in fact exactly the
same as the continuum limit of the expression (4.9) for classical
Brownian motion (with, of course, the classical phase space
distribution function replaced by the Wigner function).
To prove this assertion, consider first the case of unrestricted
propagation. Denote the path integral occuring in (4.37) by
$ \tilde K $, so
$$
\tilde K ( \dot X_f, X_0, \tau | \dot X_0, X_0, 0 )
= \int {\cal D} X \ \exp \left( - { m \over 8 \gamma k T}
\int_0^\tau dt \ddot X^2 \right) 
\eqno(4.38)
$$
where the integral is over all paths $X(t)$ satisfying
$ X(0) = X_0 $, $ \dot X (0) = X_0 $,
$ X( \tau) = X_f $, $ \dot X (\tau ) = X_f $.
This path integral is readily evaluated. The integral
is dominated by paths satisfying $ d^4 X / dt^4 = 0 $
and the above boundary conditions. These paths may be written
$$
\eqalignno{
X(t) = & X_0 + t \dot X_0 + { ( X_f - X_0 - \tau \dot X_0 ) \over
\tau^2 } t^2
\cr
& + \left( { ( \dot X_f - \dot X_0 ) \over \tau^2 }
- 2 { ( X_f - X_0 - \tau \dot X_0 ) \over \tau^3 } \right)
t^2 ( t - \tau )
&(4.39) \cr }
$$
Inserting this in the exponent and evaluating, it is readily shown
that $\tilde K $ is in fact exactly the same as the Fokker-Planck
propagator (4.5), with $p= m \dot X_f $, $ x=X_f $, $ p_0 = m \dot
X_0 $, $ x_0 = X_0 $. Therefore (4.38) is a path integral
representation of the Fokker-Planck propagator, in the unrestricted
case [\cite{Kle}].

In the restricted case, the restricted path integral (4.37)  may be
written as  a composition of propagators over a large number of
successive small time intervals, with $X(t)$ integrated over a
positive range on each time slice. However, in the limit that the
small time intervals go to zero, the {\it unrestricted} propagator
may be used to describe the propagation between neighbouring slices.
In this way we see that the restricted path integral (4.37) for the
Wigner function (4.37) coincides with our previous expression (4.9)
for the classical phase space distibution.

What we have shown may therefore be summarized as follows. We have
assumed that the parameters of the model are such that the factor $
2 m \gamma k T / \hbar^2 $ is very large. This ensures decoherence
of histories and/or density matrices. It also allows us to
approximately evaluate the integral over $\xi$ in the path integral
(4.36), leading to the expression (4.37). It then follows that, in
this approximation, the probabilities for not crossing and for
crossing $ x = 0 $ are given by the expressions (4.20), (4.21), with
the classical phase space distribution function $w_0$ replaced by
the initial Wigner function $W_0$ in the quantum case. This is the
main result of this Section.

\subhead{\bf 4(C). Properties of the Solution}

The properties of the expressions (4.20), (4.21) are not readily
seen because of the not form (4.17)--(4.19) of the restricted
propagator is not particularly transparent. However, some simple
properties of our results may be seen by examining the path integral
form (4.37) or (4.38).

It is of interest to consider the motion of a wavepacket. That is,
we take an initial state consisting of a wavepacket concentrated at
some $ x> 0 $, and moving towards the origin. We are interested in
the probability of whether it will cross $x=0$ or not during some
time interval, under the evolution by the path integral (4.37) or
(4.38).

For large $ m / ( \gamma k T ) $, which we assume, the integrand
will be strongly peaked about the unique path for which $\ddot X =
0$ with the prescribed values of $X_0$ and $ \dot X_0 $. This is of
course the classical path with the prescribed initial data. If this
path does not cross or come close to  $x=0$ during the time
interval, it will lie within the integration range $X>0$, and the
propagation is essentially the same as unrestricted propagation,
since the dominant contribution to the integral comes from the
region $ X>0 $. It is then easy to see, from the normalization of
the Wigner function, that the probability of not crossing is
approximately $1$, the intuitively expected result. 

If the classical path crosses $x=0$ during the time interval, it
will lie outside the integration range of $X(t)$ for time slices
after the time at which it crossed. The functional integration will
then sample only the exponentially small tail of the integrand, so
$W_{rr}$ will be very small. The probability of not crossing will
therefore be close to zero, again the intuitively expected result.

These results are, as stated, intuitively expected, but it is of
interest to contrast them with the unitary case described in Section
3, which has a slightly surprising feature. Consider again,
therefore, a wavepacket that starts at  $x_0 > 0 $ moving towards
the origin.  The amplitude for not crossing is given by the
restricted amplitude (3.4) and the restricted propagator (3.5).
However, in the case where the centre of the wavepacket reaches the
origin during the time interval, it is easily seen from the
propagator (3.5) that after hitting the origin there is a piece of
the wave packet which is reflected back into $x>0$ (this is the
image wavepacket that has come from $x<0$). This means that we have
the counterintuitive result that the probability for remaining in
$x>0$ is not in fact close to zero [\cite{Har,Ya1}].

Although counterintuitive, it is not particularly disturbing, since
with this initial state, the histories for crossing and not crossing
do not satisfy the consistency condition (2.6), so we should not
expect them to agree with our physical intuition. Furthermore, as we
have just shown, intuitively sensible results are obtained when the
particle is coupled to an environment to produce decoherence. In
particularly, there is no reflection of wavepackets off the origin.

The non-unitary case also gives sensible results in the case of an
initial state consisting of a superposition of wavepackets.
For example, let the initial state be of the form
$$
| \psi \ra = \a | \psi_1 \ra + \b | \psi_2 \ra
\eqno(4.40)
$$
where $ |\psi_1 \ra $ is a wavepacket concentrated at some point in
$ x>0 $ heading towards the origin, and $| \psi_2 \ra $ is also
concentrated in $ x>0 $ but is heading away from the origin.
The Wigner function of this state has the form
$$
W (p,x) = | \a |^2 W_1 (p,x) + | \b |^2 W_2 (p,x) +
{\rm interference} \ \ {\rm terms}
\eqno(4.41)
$$
where $W_1$, $W_2$ are the Wigner functions of $|\psi_1 \ra$,
$ | \psi_2 \ra $.
On inserting this in (4.38), we find the following. Firstly,
the interference terms are strongly suppressed (this is a well-known
property of evolution according to Eq.(4.1)). Secondly, using the
above results on a single wavepacket, it is easy to see that the
probabilities for crossing and not crossing are $ |\a |^2 $ and
$ | \b |^2 $ resepectively, again the expected results.

In the above simple examples, the crossing probabilities are
independent of the details of the environment, at least
approximately. It is clear that more generally, the crossing
probabilities will in fact depend on the features of the environment
({\it e.g.}, its temperature). One might find this slightly
unsettling, at least in comparison to quantum-mechanical
probabilities at a fixed moment of time, which depend only on the
state at that time and not on the details of who the property in
question might be measured. This possible dependence on the
decoherence mechanism, however, is in keeping with the point made by
Landauer and mentioned in the Introduction -- that to specify time
in quantum mechanics one has to specify the physical mechanism by
which it is measured. Furthermore, one can then expect that the
results obtained might depend to some degree on the choice of mechanism.

\def\l{{\lambda}}
\def\Tr{{\rm Tr}}
\def\a{{\alpha}}

\head {\bf 5. Histories of Crossing Densities}

We now consider a very different type of modification of the
original situation of Section 3, which leads to decoherence, and
hence to the assignment of probabilities for histories which cross
or do not cross $x=0$. We consider a system of $N$ non-interacting
free particles, and consider histories of imprecisely specified
values of number density. That is, we ask for the probability that
between $ n - \Delta n $ and $ n + \Delta n $ particles cross $x=0$
during the time interval $[0,t]$, for  $0 \le n \le N $, and $\Delta
n$ typically much smaller than $n$. As we shall see, such histories
are generally decoherent, essentially as a result of large $N$
statistics. This modification was inspired by the results of
Ref.[\cite{BrH}] on hydrodynamic histories, in which a similar
feature was observed.

We first summarize the one-particle case. Let $C$ be the class
operator for histories crossing $x=0$ during the time interval
$ [0,t]$, and $\bar C$ the
class operator for not crossing, so $ C + \bar C = 1$.
The (candidate) probabilities for crossing and
not crossing are
$$
p  = \Tr \left( C \rho C^{\dag} \right), \quad
\bar p = \Tr \left( \bar C \rho \bar C^{\dag} \right)
\eqno(5.1)
$$
respectively, and the off-diagonal term of the decoherence functional is,
$$
D = \Tr \left( C \rho \bar C^{\dag} \right)
\eqno(5.2)
$$
These quantities satisfy the relation
$$
p + \bar p + 2 {\rm Re} D = 1
\eqno(5.3)
$$

Consider the two particle case. There are three class operators,
corresponding to zero, one or two particles crossing $x=0$
during the time interval $[0,t]$. These are given by, respectively,
$$
\eqalignno{
C_0 &=  \bar C \otimes \bar C
&(5.4)\cr
C_1 &=  \bar C \otimes C + C \otimes \bar C
&(5.5) \cr
C_2 &= C \otimes C
&(5.6)\cr }
$$
and clearly $C_0 + C_1 + C_2 = 1 $.
The expressions for the case of three or more particles rapidly
become complicated, but we are saved by a useful trick, used in
Ref.[\cite{BrH}] (and similar to a trick used in studies of 
random walks [\cite{MoW}]).
In the $N$ particle case, the class operator corresponding to $n$
particles crossing is given by
$$
C_n = { 1 \over 2 \pi} \int_{- \pi}^{\pi} d \lambda
\ e^{-i \lambda n} 
\ \left( \bar C + e^{i \lambda} C \right) \otimes
\left( \bar C + e^{i \lambda} C \right) \otimes
\ \cdots
\eqno(5.7)
$$
where there are $N$ terms in the tensor product. How this expression
works is that in tensor product terms, the coefficient of
$ e^{-i \l n} $ consists of all possible combinations of terms
consisting of $n$ $C$'s and $(N-n)$ $\bar C$'s.
Eventually we will be interested in a coarse graining
over $n$, which consists of binning $n$ into ranges of
width $ 2 \Delta n$, labeled by $\bar n$,
$$
C_{\bar n} = \sum_{n \in \bar n} \ C_n
\eqno(5.8)
$$
We will not carry this out explicitly, since the result of
doing this is intuitively clear. Explicit coarse grainings
of this type in a related problem were carried out in Ref.[\cite{BrH}].

The decoherence functional for histories of precisely specified
values of $n$ is
$$
D(n,n') = \Tr \left( C_n \ \rho \otimes \rho \otimes \cdots 
\otimes \rho
\ C_{n'}^{\dag} \right)
\eqno(5.9)
$$
where we have assumed a factored initial state for the
$N$ particle system. Inserting the above expression for $C_n$,
this may be written
$$
D(n,n') = { 1 \over (2 \pi )^2 }
\ \int_{- \pi}^{\pi} d \lambda
\ \int_{- \pi}^{\pi} d \lambda'
\ e^{ -i \lambda n + i \lambda' n'}
\ \left( e^{i (\l -\l')} p + e^{i \l} D
+ e^{-i \l'} D^* + \bar p \right)^N
\eqno(5.10)
$$
Using the binomial expansion to expand the integrand,
the integral over $\l$ may be carried out, with the
result,
$$
D(n,n') = {1 \over 2 \pi} \left( {N \atop n} \right)
\ \int_{-\pi}^{\pi} d \l' \ e^{i \l' n'}
\ \left( \bar p + e^{-i \l'} D^* \right)^{N-n}
\ \left( D + e^{-i \l' } p \right)^n
\eqno(5.11)
$$
Further use of the binomial theorem permits the
remaining integral to be done, with the result
$$
D(n,n') = \left( {N \atop n} \right) {\bar p}^{N-n-n'}
(D^*)^{n'} D^n \ \sum_{k=0}^n \left( { N-n \atop n' -k} \right)
\left( { n \atop k} \right) \ \left( { p \bar p \over
|D|^2 } \right)^k
\eqno(5.12)
$$
for $ n \le n' $. For $ n \ge n'$, on the other hand, one obtains,
$$
D(n,n') = \left( {N \atop n} \right) {\bar p}^{N-n}
p^{n'} D^{n-n'} \ \sum_{k=o}^{N-n} \left( { N-n \atop k} \right)
\left( { n \atop n'- k} \right) \ \left( { |D|^2 \over p \bar p }
\right)^k
\eqno(5.13)
$$

It is useful in (5.11) to rewrite the integral as a complex contour
integral. Let $z= e^{-i \l'} $. Then we obtain
$$
D(n,n') = { 1 \over 2 \pi i} 
\left( { N \atop n} \right) \int {dz \over z^{n' +1} }
\ \left( \bar p + D^* z \right)^{N-n}
\ \left( D + p z \right)^n
\eqno(5.14)
$$
where the integral is along any closed contour about the origin. Now
performing the rescaling $ z \rightarrow ( \bar p / D^* ) z $,
this becomes
$$
D(n,n') = {1 \over 2 \pi i} \left( {N \atop n} \right) {\bar p}^{N-n-n'}
(D^*)^{n'} D^n 
\ \int dz \ z^{-n'-1} \ ( 1 + z)^{N-n} \ ( 1 + \a z )^n
\eqno(5.15)
$$
where $ \a = p \bar p / |D|^2 $.

The discrete sums in (5.12) and (5.13) can be evaluated in terms
of a hypergeometric function, $F$. For example, (5.12) yields
$$
D(n,n') = \left( {N \atop n} \right) 
\ \left( { N-n \atop n' } \right)
\ {\bar p}^{N-n-n'} (D^*)^{n'} D^n 
\ F \left( -n, -n'; N - n -n'+1, \alpha \right)
\eqno(5.16)
$$
However, the hypergeometric function is of the degenerate type
(and can be written as a finite hypergeometric series)
for which asymptotic forms are not easily found, although this
exact expression may be of use for computer plots. We will instead
therefore consider asymptotic forms of the expressions (5.12),
(5.13) and (5.15).

Consider first the case of very large 
$\alpha$. This is the case in which there is some degree of decoherence
of the one particle system, but perhaps not sufficient to assign
probabilities defined to satisfactory precision.
We shall see that this is exponentially enhanced in the $N$ particle case. 

Taking $N, n, n'$ to be of the same order
(although not necessarily large), for $ \alpha >> N^2 $, the discrete
sum (5.12) is dominated by the $k=n$ term, and we find
$$
D(n,n') =
{ N! \over n! (N-n')! (n'-n)!}
\ {\bar p}^{N-n-n'} (D^*)^{n'} D^n \ \alpha^n 
\ \left( 1 + O \left( N^2  \over \alpha \right)  \right)
\eqno(5.17)
$$
A reasonable measure of approximate decoherence is the size of the
decoherence functional in comparison to its diagonal terms.
Here, this is given by
$$
\e \ \equiv \ { | D(n,n') |^2 \over D(n,n) D(n',n') }
\ \approx
\ {1 \over \alpha^{n' -n} }
\ { n'! (N-n)! \over n! (N-n')! ( (n'-n)! )^2 }
\eqno(5.18)
$$
Since $\alpha >> N^2 $, the dominant term is the term depending
on $\alpha$. For $n'-n$ reasonably large (recall that this is the
case $n' > n $), the degree of decoherence of the N particle
case is exponentially enhanced compared to the one particle case.

Of course, $n'$ and $n$ may differ by a small number, like $1$
or $2$, in which case the degree of decoherence would then not be
very good. The point is, however, that we are envisaging the further
coarse graining (5.8). As can be seen from similar calculations
in Ref.[\cite{BrH}], this would have the effect of replacing $ n$ and
$n'$ by coarse grained variables $\bar n$ and ${\bar n}'$. These
can differ by no less than the coarse graining parameter $2 \Delta n$,
which is taken to  be large. The degree of decoherence is therefore
of order $ \alpha^{-2 \Delta n}$, which will be very small.

Given decoherence for the case of large $\alpha$, we may now assign
probabilities. These are given by
$$
p(n) = \left( { N \atop n} \right) p^n {\bar p}^{N-n}
\left( 1 + { (N-n) n \over \alpha} + \cdots \right)
\eqno(5.19)
$$
For large $N,n$, this becomes, to leading order,
$$
p(n) \ \sim \ \exp \left( - { N \over 2 n (N-n) }
\left( n - { p N \over ( p + \bar p ) } \right)^2 \right)
\eqno(5.20)
$$
Inserting the most probable value of $n$ in the width,
this becomes
$$
p(n) \ \sim \ \exp \left(
- { N ( p + \bar p )^2 \over 2 p \bar p }
\left( { n \over N} - { p \over ( p +\bar p)} \right)^2 \right)
\eqno(5.21)
$$
Note that we cannot take $p + \bar p = 1 $ since these are
not consistent probabilities.

This is a gratifying result. It shows that the relative frequency
with which the particles cross is strongly peaked about the
value $ p / ( p + \bar p ) $. 
Also notice that
$$
\left( { n \over N} - { p \over ( p +\bar p)} \right)^2
= \left( { N-n \over N} - { \bar p \over ( p + \bar p )} \right)^2
\eqno(5.22)
$$
which is consistent with the notion that the relative
frequency of not crossing is $ \bar p / ( p + \bar p ) $.
These results are tantamount to taking the probabilities for
crossing and not crossing in the single particle case
to be not $p$ and $\bar p$, but
$ p / ( p + \bar p ) $ and $ \bar p / ( p + \bar p ) $
(which clearly add to $1$, as required).

Again we should be considering coarse grained values of $n$
but it is clear that this will effect only the width of the peak
and not the configurations about which the distribution is peaked.

Another case which is amenable to straightforward analysis is
the case $\alpha = 1$. This might not be exactly reachable in practice,
but it represents the extreme case in which the decoherence of the
one particle case is a bad as it can possibly get. From either
(5.12) or (5.15), we find
$$
D(n,n') = \left( { N \atop n } \right) \left( { N \atop n'} \right)
\ {\bar p}^{N-n-n'} (D^*)^{n'} D^n
\eqno(5.23)
$$
It is straightforward to show that in this case
$$
| D(n,n') |^2 = D(n,n) D(n',n')
\eqno(5.24)
$$
hence the decoherence in the $N$ particle case is just
as bad as the one particle case!

Now we consider the somewhat harder and more general 
case of $\alpha >1 $ but not arbitrarily large. Here
we resort to some more sophisticated techniques to
expand the contour integral (5.15) in the limit of
large $N,n,n'$.

The integral (5.15) may be written
$$
D(n,n') = \left( {N \atop n} \right) {\bar p}^{N-n-n'}
(D^*)^{n'} D^n \ J
\eqno(5.25)
$$
where
$$
J = { 1 \over 2 \pi i} 
\ \int dz \ z^{-n'-1} \ \left( f(z) \right)^N
\eqno(5.26)
$$
and
$$
f(z) = (1 + z )^{1-n/N} \ (1 + \alpha z )^{n/N}
\eqno(5.27)
$$
This integral, for large $N$, has the asymptotic form
$$
J  \ \sim \ { \left( f (\rho) \right)^N \over
\left( 2 \pi N \kappa_2 (\rho) \right )^{\half} \rho^{n'} }
\ \left( 1 + O \left( {1 \over N} \right) \right)
\eqno(5.28)
$$
Here, $\rho $ is the unique positive solution to the equation,
$$
N \rho { f'( \rho ) \over f(\rho ) } = n'
\eqno(5.29)
$$
which, in this case, reads,
$$
\alpha \left( N - n' \right) \rho^2
+ \left( N - n - n' + \alpha (n-n') \right) \rho - n' = 0
\eqno(5.30)
$$
and
$$
\kappa_2 (\rho) = \rho { f'( \rho ) \over f( \rho ) }
+ \rho^2 \left(
{ f^{\prime\prime} (\rho ) \over f(\rho ) }
- \left( { f'(\rho) \over f( \rho ) } \right)^2 
\right) 
\eqno(5.31)
$$

The origin of this formula is as follows [\cite{Ego,Goo}].
The integration contour
in (5.26) is any closed contour about the origin. 
Let $z = \rho e^{i \theta} $, where $\rho$ is arbitrary.
The idea is to take 
a circular contour whose radius is chosen in such a way that the
dominant contribution to the integral for large $N$ comes
from the immediate neighbourhood of $\theta = 0$.
In terms of $\rho$ and $\theta$ the integal becomes
$$
J = {1 \over 2 \pi \rho^{n'} } \int_{-\pi}^{\pi} d \theta
\ e^{-in' \theta} \ \left( f (\rho e^{i \theta} ) \right)^N
\eqno(5.32)
$$
Now expand the integrand about $\theta = 0 $. We have
$$
\eqalignno{
\left( f (\rho e^{i \theta} ) \right)^N
&= \exp \left( N \ln f (\rho e^{i \theta} ) \right)
\cr
&= \left( f(\rho ) \right)^N
\ \exp \left( i N \theta \rho { f'( \rho ) \over f ( \rho ) }
- \half N \theta^2 \kappa_2 (\rho) + O ( N \theta^3 )  \right)
&(5.33) \cr }
$$
where $\kappa_2$ is given by (5.32). Now clearly if $\rho$,
which is so far arbitrary, is chosen to satisfy (5.29), the
linear term in the exponent in the whole integrand vanishes.
For large $N$ the integral over $\theta$ is then a Gaussian strongly
concentrated around $\theta = 0 $, and 
may be done with the desired result (5.28).

(Note that it was not necessary to use this more elaborate
asymptotic expansion technique in Ref.[\cite{BrH}]. There, the
integral analagous to Eq.(5.10) has the property that the modulus of
the integrand is less than $1$ and equal to $1$ when the $\lambda$
parameters are zero, so it was possible to evaluate for large $N$ by
expanding about zero. Here, the norm of the integrand (5.10) does
not have this property).

The decoherence functional is therefore given by (5.25)
with, to leading order,
$$
J = J_{n n'} = \left( 1 + \rho_{n n'} \right)^{N-n}
\ \left( 1 + \alpha \rho_{n n'} \right)^n
\ \rho_{n n'}^{-n'}
\eqno(5.34)
$$
and 
$$
\rho_{n n'} = {
- N + n + n' - \alpha ( n - n' )
+ \left[ \left( N - n - n' + \alpha ( n - n' ) \right)^2
+ 4 \alpha n' ( N - n' ) \right]^{\half} \over
2 \alpha ( N - n' ) }
\eqno(5.35)
$$
The candidate probabilities for the histories are 
$$
p(n) = \left( {N \atop n} \right) {\bar p}^{N-2n}
\bigl | D \bigr |^{2n}
\ \left( 1 + \rho_n \right)^{N-n} \left( 1 + \alpha \rho_n \right)^n
\rho_n^{-n}
\eqno(5.36)
$$
where
$$
\rho_n =
{
- N + 2n
+ \left[ \left( N - 2n \right)^2
+ 4 \alpha n ( N - n ) \right]^{\half} \over
2 \alpha ( N - n ) }
\eqno(5.37)
$$
The probabilities may be assigned when the degree of decoherence,
$$
\e = { | D(n,n') |^2 \over D(n,n) D(n',n') }
\ = 
\ { n'! (N-n')! \over n! (N-n)! } 
\ { | J_{nn'} |^2 \over J_{nn} J_{n'n'} }
\eqno(5.38)
$$
is small. Eqs.(5.34)--(5.37) give the degree of decoherence and the
expressions for the probabilities for all values of $\a$ 
when $N$, $n$, $n'$ are large. Since these are not very transparent,
it is useful to examine them in more detail for special cases.

Above we extracted the leading order for very large $\a$
(essentially $ \a > > N^2 $). We may now improve on this by
expanding (5.34)--(5.37) for the case $ \a >> 1 $, if we also assume
that $ |n'-n| $ is about the same order of magnitude as $N,n,n'$.
A straightforward but tedious calculation shows that 
the degree of decoherence is
$$
\epsilon \sim \a^{-|n-n'|}
\eqno(5.39)
$$
to leading order, which will be very small.
Furthermore, the probabilities are given by
(5.21). Hence the result obtained for the case $\a >> N^2 $ above
also hold for $ \a >> 1 $.

Another case easily handled is the case $ \a = 1 + \delta $,
where $0 < \delta << 1 $. 
Recall that
for $\a = 1$ there is no decoherence (Eq.(5.24)), so it is
interesting to see how large $\a$ needs to be before decoherence is
achieved. Again a straightforward calculation shows that, to leading
order, the degree of decoherence is
$$
\e \sim \ \exp \left( - { (n-n')^2 \over N } \delta \right)
\eqno(5.40)
$$
Assuming again that $n-n'$ and $N$ are of about the same order,
approximate decoherence is achieved if $ \delta >> 1/|n-n'| $. Hence
$\a$ does not have to be very much greater than $1$ in order to
achieve approximate decoherence. The probabilities in this
case are, to leading order,
$$
p(n) \sim  \left( { N \atop n} \right)^2 {\bar p}^{N-n} p^n \ \exp
\left( \delta { n^2 \over N } \right)
\eqno(5.41)
$$
For large $N$, $n$, and recalling that $\delta < < 1 $,
this has the asympotic form,
$$
p(n) \ \sim \ \exp \left( - { N \over  n (N-n) }
\left( n - { p^{\half} N \over ( p^{\half} + \bar p^{\half} ) } 
\right)^2 \right)
\eqno(5.42)
$$
(This is easily seen by noting that (5.41) is the square of the
leading order term in (5.19) with $p$, $\bar p$ replaced by
$ p^{\half} $, ${\bar p}^{\half} $.) This case therefore
corresponds to regarding the expressions 
$ p^{\half}/ (p^{\half} + {\bar p}^{\half} ) $ and
$ {\bar p}^{\half}/ (p^{\half} + {\bar p}^{\half} ) $ as the
probabilities for crossing and not crossing in the one particle case.

Finally, we note that all of the analysis of this section does not
in fact specifically concern the crossing time problem. It would
apply to any situation in which the original system consists of a
coarse graining into just two histories, the system is replicated
$N$ times, and projections onto the relative frequency $ f = n / N
$, suitably coarse grained, are considered.  This is not unrelated
to the Finkelstein--Graham--Hartle theorem [\cite{FGH}], which shows
that the conventional probabilistic interpretation of quantum theory
can arise from consideration of the eigenstates of relative
frequency operator of the entire closed system. Here, we have shown
that the relative frequency for histories is typically decoherent
for large $N$ (in this connection, see also Ref.[\cite{Ohk}]).

\head {\bf 6. Summary and Conclusions}

For the closed system consisting of a single point particle in
non-relativistic quantum mechanics, probabilities generally cannot
be assigned to histories partitioned according to whether or not
they cross $x=0$ during a fixed time interval. We have shown in this
paper, however, that by making modifications to this basic physical
situation, decoherence may be achieved and probabilities assigned
for arbitrary initial states.

The first modification we considered was to couple the particle to a
thermal environment. This corresponds to continuous imprecise
measurements of the particle's position. The desired probabilities
are given by Eqs.(4.20), (4.21), where $w_0$ is taken to be the
initial Wigner function.

The second modification consisted of replicating the system $N$
times, and then considering the number density of particles crossing
$x=0$ in the limit of large $N$. This less obviously corresponds to
a particular type of measurement, but on general grounds, since
there is decoherence (rather than just consistency), there is a
correspondence with some kind of measurement (although not
necessarily a physically realizable one). The probabilities in a
regime of interest are given by Eq.(5.21).

In each case, when decoherence is achieved, the resultant
probabilities depend, at least to some degree, on the mechanism
producing decoherence, and this is to be expected.

\head{\bf Acknowledgements}

We are grateful to Lajos Di\'osi, Chris Isham, Jim Hartle, Carlo
Rovelli and Jason Twamley for useful conversations.  J.H. is
grateful for the hospitality of the Schr\"odinger Institute, Vienna,
at which part of this work was carried out. E.Z. was supported in
part by the A.S.Onassis Public Benefit Foundation.

\references

\def\pr{{\sl Phys. Rev.\ }}
\def\prl{{\sl Phys. Rev. Lett.\ }}
\def\prep{{\sl Phys. Rep.\ }}
\def\jmp{{\sl J. Math. Phys.\ }}
\def\rmp{{\sl Rev. Mod. Phys.\ }}

\def\np{{\sl Nucl. Phys.\ }}

\def\annp{{\sl Ann. Phys. (N.Y.)\ }}

\def\amjp{{\sl Am. J. Phys.\ }}

\refis{AuK} A.Auerbach and S.Kivelson, \np {\bf B257}, 799 (1985).
% The path decomposition expansion and multidimensional tunneling

\refis{vB} P. van Baal, ``Tunneling and the path decomposition expansion'',
in {\it Lectures on Path Integration: Trieste 1991}, edited by
H.A.Cerdeira {\it et al.} (World Scientific, Singapore, 1993).

\refis{BaJ} N.Balazs and B.K.Jennings, \prep {\bf 104}, 347 (1984),
M.Hillery, R.F.O'Connell, M.O.Scully and E.P.Wigner, \prep {\bf
106}, 121 (1984); 
V.I.Tatarskii, {\sl Sov.Phys.Usp} {\bf 26}, 311 (1983). 

\refis{BuT} M.A.Burschka and U.M.Titulaer,
{\sl J.Stat.Phys.} {\bf 25}, 569 (1981); {\bf 26}, 59 (1981);
{\sl Physica} {\bf 112A}, 315 (1982).

\refis{BoD} A.Boutet de Monvel and P.Dita, {\sl J.Phys.}
{\bf A23}, L895 (1990).
% Analytic solution for a boundary value problem in the theory of
% Brownian motion.

\refis{BrH} T.Brun and J.J.Halliwell, {\sl Phys.Rev.} {\bf D54}, 2899 (1996).
% Decoherence of Hydrodynamic Histories: A Simple Spin Model.

\refis{CaL} A.O.Caldeira and A.J.Leggett, {\sl Physica} {\bf
121A}, 587 (1983).

\refis{Car} H.S.Carslaw, {\sl Proc.Lond.Math.Soc} {\bf 30}, 121 (1899).

\refis{DoH} H.F.Dowker and J.J.Halliwell, {\sl Phys. Rev.} {\bf
D46}, 1580 (1992).

\refis{Ego} G.P.Egorychev, {\it Integral Representation and the
Computation of Combinatorial Sums}, Translations of Mathematical
Monographs, Vol.59 (American Mathematical Society, Providence, RI,
1984).

\refis{FeV} R.P.Feynman and F.L.Vernon, \annp {\bf 24}, 118 (1963).

\refis{Fer} H.Fertig, \prl {\bf 65}, 2321 (1990).
% Traversal-time distribution and the uncertainty principle in
% quantum tunneling

\refis{FGH} D.Finkelstein, {\sl Trans.N.Y.Acad.Sci.} {\bf 25}, 
621 (1963);
%{\it The Logic of Quantum Physics.} 
N.Graham, in {\it The Many Worlds Interpretation of
Quantum Mechanics}, B.S.DeWitt and N.Graham (eds.) (Princeton
University Press, Princeton, 1973);
%{\it The Measurement of Relative Frequency.} 
J.B.Hartle, \amjp {\bf 36}, 704 (1968).
%{\it Quantum Mechanics of Individual Systems.} 
See also, E.Farhi, J.Goldstone and S.Gutmann,  \annp {\bf 192}, 
368 (1989).
%{\it How Probability Arises in Quantum Mechanics.} 

\refis{GeH1} M.Gell-Mann and J.B.Hartle, 
in {\it Complexity, Entropy 
and the Physics of Information, SFI Studies in the Sciences of Complexity},
Vol. VIII, W. Zurek (ed.) (Addison Wesley, Reading, 1990); and in
{\it Proceedings of the Third International Symposium on the Foundations of
Quantum Mechanics in the Light of New Technology}, S. Kobayashi, H. Ezawa,
Y. Murayama and S. Nomura (eds.) (Physical Society of Japan, Tokyo, 1990).

\refis{GeH2} M.Gell-Mann and J.B.Hartle,
{\sl Phys.Rev.} {\bf D47}, 3345 (1993).

\refis{GeH4} M.Gell-Mann and J.B.Hartle, 
in {\it Proceedings of the 4th Drexel Symposium on Quantum
Non-Integrability --- The Quantum-Classical Correspondence} edited 
by D.H.Feng and B.L.Hu
(International Press, Boston/Hong-Kong, 1996)

\refis{Goo} I.J.Good,
{\sl Ann.Math.Stat.} {\bf 28}, 861 (1957);
% Saddlepoint methods for the multinomial distribution
{\bf 32}, 535 (1961).
% The multivariate saddlepoint method and chi-squred for the
% multinomial distribution

\refis{Gri} R.B.Griffiths, {\sl J.Stat.Phys.} {\bf 36}, 219 (1984);
{\sl Phys.Rev.Lett.} {\bf 70}, 2201 (1993);
{\sl Am.J.Phys.} {\bf 55}, 11 (1987).

\refis{GRT} N.Grot, C.Rovelli and R.S.Tate, preprint
quant-ph/9603021 (1996).
% Time--of--arrival in quantum mechanics.

\refis{Hal1} J.J.Halliwell, ``Aspects of the Decoherent Histories
Approach to Quantum Theory'',
in {\it Stochastic Evolution of Quantum States in Open Systems and 
Measurement Processes}, edited by
L.Di\'osi, L. and B.Luk\'acs (World Scientific, Singapore, 1994).

\refis{Hal2} J.J.Halliwell,
``A Review of the Decoherent Histories Approach to Quantum Mechanics'',
in {\it Fundamental Problems in Quantum Theory}, 
edited by D.Greenberger and A.Zeilinger,
Annals of the New York Academy of Sciences, Vol 775, 726 (1994).

\refis{Hal3} J.J.Halliwell, {\sl Phys.Lett} {\bf A207}, 237 (1995).
% An operator derivation of the path decomposition expansion.

\refis{Hal4} J.J.Halliwell, in {\it General Relativity and Gravitation
1992}, edited by R. J. Gleiser, C. N. Kozameh and O. M. Moreschi
(IOP Publishers, Bristol, 1993).

\refis{HaO} J.J.Halliwell and M.E.Ortiz, {\sl Phys.Rev.} 
{\bf D48}, 748 (1993).

\refis{HaZ} J.J.Halliwell and A.Zoupas,
{\sl Phys.Rev.} {\bf D52}, 7294 (1995);
% ``Quantum State Diffusion, Density Matrix Diagonalization
% and Decoherent Histories: A Model''.
``Post-decoherence density matrix propagator for quantum 
Brownian motion'',
IC preprint 95-96/67, quant-ph/9608046 (1996),
accepted for publication in {\sl Phys.Rev.D} (1997).

\refis{Har} J.B.Hartle, {\sl Phys.Rev.} {\bf D44}, 3173 (1991).
%{\it Spacetime Coarse-Grainings in Non-Relativistic Quantum Mechanics.}

\refis{Har0} J.B.Hartle, {\sl Phys.Rev.} {\bf D38}, 2985 (1988).
% Quantum kinematics of spacetime. II. 
% A model quantum cosmology with real clocks.

\refis{Har1} J.B.Hartle, in {\it Quantum Cosmology and Baby
Universes}, S. Coleman, J. Hartle, T. Piran and S. Weinberg (eds.)
(World Scientific, Singapore, 1991).  
%{\it The Quantum Mechanics of
%Cosmology.}

\refis{Har2} J.B.Hartle, {\sl Phys.Rev.} {\bf D37}, 2818 (1988).

\refis{Har3} J.B.Hartle, in {\it Proceedings of the 1992 Les Houches Summer
School, Gravitation et Quantifications}, edited by B.Julia and J.Zinn-Justin 
(Elsevier Science B.V., 1995)
%{\it Spacetime Quantum Mechanics and the Quantum Mechanics of Spacetime.}

\refis{HaS}  E.H.Hauge and J.A.Stovneng, \rmp {\bf 61}, 917
(1989).
% Tunneling times: a critical review.

\refis{Hol} A.S.Holevo, {\it Probabilistic and Statistical Aspects
of Quantum Theory} (Publisher?, 1982). Pages 130--197. 

\refis{Ish} C. Isham, \jmp {\bf 23}, 2157 (1994);
%{\it Quantum Logic and the Histories Approach to Quantum Theory.}
C. Isham and N. Linden, \jmp {\bf 35}, 5452 (1994);
{\bf 36}, 5392 (1995);
%{\it Quantum Temporal Logic and Decoherence Functionals in the
%Histories Approach to Generalized Quantum Theory}.
%"Continuous histories and the history group in generalised quantum theory",
%(with N. Linden).
C. Isham, N. Linden and S.Schreckenberg,
\jmp {\bf 35}, 6360 (1994).
%{\it The Classification of Decoherence Functionals:
%An Analogue of Gleason's Theorem}.

\refis{JoZ} E.Joos and H.D.Zeh, {\sl Zeit.Phys.} {\bf B59}, 223
(1985).
%{\it The Emergence of Classical Properties through Interaction with
%an Environment.}

\refis{Kle} That the propagator for the Fokker-Planck equation can
be represented in terms of a configuration space path integral of
this form does not appear to be widely known. See, however,
H.Kleinert, {\it Path Integrals in Quantum Mechanics, Statistics and
Polymer Physics} (World Scientific, Singapore, 1990), pages
635--644.

\refis{KoA} D.H.Kobe and V.C.Aguilera--Navarro, 
\pr {\bf A50}, 933 (1994).
% Derivation of the energy-time uncertainty relations.

\refis{Kum} N.Kumar, {\sl Pramana J.Phys.} {\bf 25}, 363 (1985).
% Quantum First Passage Problem.

\refis{Lan} R.Landauer {\sl Rev.Mod.Phys.} {\bf 66}, 217 (1994);
% Barrier interaction time in tunneling.
{\sl Ber.Bunsenges.Phys.Chem} {\bf 95}, 404 (1991).
% Traversal time in tunneling

\refis{MaT} L.Mandelstamm and I.Tamm, {\sl J.Phys.} {\bf 9}, 249
(1945).
% The uncertainty relation between energy and time in
% non-relativistic quantum mechanics.

\refis{MaW} T.W.Marshall and E.J.Watson, {\sl J.Phys.} {\bf A18},
3531 (1985);
% A drop of ink falls from my pen...It comes to earth, I know not when
{\bf A20}, 1345 (1987).
% The analytic solutions of some boundary layer problems in the
% theory of Brownian motion

\refis{MiH} R.J.Micanek and J.B.Hartle, {\sl Phys.Rev.} {\bf A54},
3795 (1996). 
% Santa Barbara preprint
% NSF-ITP-96-13, quant-ph/9602023 (1996).
% Nearly instantaneous alternatives in quantum mechanics.

\refis{MoW}  E.Montroll and B.West, in {\it Fluctuation
Phenomena}, edited by E.Montroll and J.Lebowitz (North Holland,
Amsterdam, 1979).

\refis{Ohk} Y.Ohkuwa, {\sl Phys.Rev.} {\bf D48}, 1781 (1993).
%Decoherence Functional and Probability Interpretation.

\refis{Omn} R.Omn\`es, {\sl J.Stat.Phys.} {\bf 53}, 893 (1988);
{\bf 53}, 933 (1988);
{\bf 53}, 957 (1988);
{\bf 57}, 357 (1989);
{\bf 62}, 841 (1991);
{\sl Ann.Phys.} {\bf 201}, 354 (1990); 
{\sl Rev.Mod.Phys.} {\bf 64}, 339 (1992);
{\it The Interpretation of Quantum Mechanics}
(Princeton University Press, Princeton, 1994).

\refis{PZ} J.P.Paz and W.H.Zurek, \pr {\bf D48}, 2728 (1993).

\refis{Per} A.Peres, {\it Quantum Theory: Concepts and Methods}
(Kluwer Academic Publishers, Dordrecht, 1993). Pages 405--417.

\refis{PHZ} J.P.Paz, S.Habib and W.Zurek, \pr {\bf D47}, 488 (1993).

\refis{ScZ} L.Schulman and R.W.Ziolkowiski, in {\sl Path integrals from
meV to MeV}, edited by V. Sa-yakanit, W. Sritrakool, J. Berananda, M. C.
Gutzwiller, A. Inomata, S. Lundqvist, J. R. Klauder and L. S. Schulman
(World Scientific, Singapore, 1989).

\refis{Sie} A.J.F.Siegert, {\sl Phys.Rev.} {\bf 81}, 617 (1951).

\refis{Time} See, for example,
E.P.Wigner, \pr {\bf 98}, 145 (1955); F.T.Smith, \pr
{\bf 118}, 349 (1960); E.Gurjoy and D.Coon, {\sl Superlattices and
Microsctructures} {\bf 5}, 305 (1989); C.Piron, in {\it
Interpretation and Foundations of Quantum Theory}, edited by
H.Newmann (Bibliographisches Institute, Mannheim, 1979);
G.R.Allcock, \annp {\bf 53}, 253 (1969);
{\bf 53}, 286 (1969); {\bf 53}, 311 (1969).

\refis{WaU} M.C.Wang and G.E.Uhlenbeck, {\sl Rev.Mod.Phys.}
{\bf 17}, 323 (1945). Reprinted in, {\it Selected Papers on Noise
and Stochastic Processes}, edited by N.Wax (Dover Publications, New
York, 1954).

\refis{Ya1} N.Yamada (unpublished).

\refis{YaT} N.Yamada and S.Takagi, {\sl Prog.Theor.Phys.}
{\bf 85}, 985 (1991); {\bf 86}, 599 (1991); {\bf 87}, 77 (1992);
N. Yamada, {\sl Sci. Rep. T\^ohoku Uni., Series 8}, {\bf 12}, 177
(1992).

\refis{Zaf} E.Zafiris, ``Stochastic Phase Space Approach to the
First Passage Time Probability Problem'', Imperial College preprint
TP/96-97/05 (1997).

\refis{Zur} W.Zurek, {\sl Prog.Theor.Phys.} {\bf 89}, 281 (1993);
{\sl Physics Today} {\bf 40}, 36 (1991);
in, {\it Physical Origins of Time Asymmetry}, edited by 
J.J.Halliwell, J.Perez-Mercader and W.Zurek (Cambridge
University Press, Cambridge, 1994).

\endreferences
\end